# MPC-Based Real-Time Charging Coordination for Electric Vehicle Aggregator to Provide Regulation Service in a Market Environment

Liling Gong, Ye Guo*
Tsinghua-Berkeley Shenzhen Institute
Tsinghua University
Shenzhen, China

Hongbin Sun
Department of Electrical Engineering
Tsinghua University
Beijing, China

*Abstract*—The optimal operation problem of electric vehicle aggregator (EVA) is considered. An EVA can participate in energy and regulation markets with its current and upcoming EVs, thus reducing its total cost of purchasing energy to fulfill EVs' charging requirements. An MPC based optimization model is developed to consider future arrival of EVs as well as energy and regulation prices. The index of CVaR is used to model risk-averseness of an EVA. Simulations on the 1000-EV test system validate the effectiveness of our work in achieving a lucrative revenue while satisfying the charging requests from EV owners.

*Index Terms*--electric vehicle aggregator, model predictive control, frequency regulation, electricity market.

## I. INTRODUCTION

The adoption rate of electric vehicles (EVs) was accelerated during last decade: the number of global EVs has been 10 million by 2020 but only 17,000 by 2010[1]. This is because EVs have an indispensable role to reach net-zero emissions. On one hand, EVs can replace combustion engine vehicles to reduce the emission of greenhouse gases. One the other hand, charging EVs is promising to provide flexibility services to the grid, including energy arbitrage and ancillary services provision.

With the increasing penetration of renewable generations, more regulation capabilities are needed to effectively manage the rapid variation of supply and demand in power system. In the near future, EVs will be one of the most realistic forms to provide frequency regulation because most EVs remain idle for almost 90% of the time[2]. Furthermore, the performance-based regulation compensation provides sufficient incentive for fast-ramping resources such as EVs to quickly and accurately respond to AGC signal. However, it is impractical for individual EV's participation due to two main reasons: i) Most electricity markets are carried out on an MW basis (the minimum regulation capability in PJM is 0.1MWh[3]); ii) It is difficult for ISO to manage such a significant number of transactions in system level. Therefore, a new entity, electric vehicle aggregator (EVA) has been proposed to coordinate the interactions between EV owners and market ISO.

Most of the literature about EVs providing regulation fall into two main categories: bidding strategies[4]-[6] and charging allocations[7]-[8]. Owing to the uncertain nature of market clearing prices and EV fleet characteristics, stochastic programming[4], chance-constrained programming[5] and robust optimization[6] have been employed to maximize EVA profits in the real-time (RT) electricity market. The works in the second category, propose RT control algorithms for EVA to allocate the charging power of EVs, whereby EVs connected are responsible for regulation provisions. In [7], a real-time greedy-index dispatching policy is proposed, which transforms the dispatch problem from a high-dimensional space into an 1-D space while preserving the solution optimality. In [8], a hierarchical V2G control strategy is modeled to allocate the regulation task within the regulation capacity of EVs, which considers both the ACE-based frequency regulation and the expected battery SoC levels of EV owners.

Note that the role of EVA is twofold: firstly, it participates in electricity markets based on its bidding strategies; then, it allocates the charging rates of EVs according to market cleared settlements. Therefore, an EVA operation problem considers both the market bidding process and charging allocation with regulation provision. A framework for charging management of an EVA is proposed in [9], and the RT allocation algorithm is developed based on linear programs with charging priority weights. In [10], a Laxity-SoC-based smart charging criteria is proposed during RT operation. However, these aforementioned heuristic dispatching methods have a negative financial impact on the future market performance. To conclude, such the RT operation problem of EVA has a challenge emerged: how to pursue the optimum market profits while simultaneously achieving the delivery of cleared regulation capacities and charging requirements from EV owners?

To deal with the coupling effect of the EV charging requests and EVA economic optimum, model predictive control (MPC) method is widely adopted to optimize multi-time decisions within additional constraints. [11] develops a RT charging controller which operates an EV fleet to provide regulation, however, the MPC scheme is only used to predict AGC signals. [12] proposes an MPC scheme for EVA participating in the regulation market with SARIMA prediction on the regulation prices, however, energy costs are ignored in the modeling. The key distinction of EVs lies in the uncertainties of the volumes and time frames of charging demands of future upcoming EVs. Compared to[11]-[12], in this paper we consider the future uncertain demands when operating RT charging rates of connected EVs.

This paper aims to develop a RT charging coordination methodology for EVA to participate in energy and regulation markets. The main contributions are summarized as follows: i) An equivalent virtual EV model is built to predict the charging demands of upcoming EVs; ii) A Conditional Value at Risk (CVaR) based two-stage stochastic programming is formulated to allocate the cleared regulation task to connected EVs and hedge the risk of bidding with uncertainties of future upcoming EVs as well as energy and regulation prices; iii) A rolling-horizon operation framework is developed to minimize the mismatch of regulation bid amounts and actual capability, while guaranteeing the charging requests from EV owners.

## II. Problem Description

We consider the RT charging problem for an EVA, where all EVs plug-in are coordinated by the aggregator. Once an EV arrives, it will declare its departure time and expected SoC to EVA, and the aggregator needs to fulfill the charging request of each individual EV before its leaving. Unidirectional interaction with the grid is adopted in this paper, where chargers are not capable to discharge energy from EVs back to the grid. However, EVs are still capable of participating in the energy and regulation markets by adjusting their charging powers away from the Preferred Operation Point (POP). As illustrated in Figure 1, the aggregator submits an offer in RT wholesale electricity market and manages a fleet of connected EVs which are responsible to follow the regulation signals. In exchange, EVA would pay lucrative compensations to EV owners for the regulation provisions. This EV compensation structure is outside the scope of this paper.

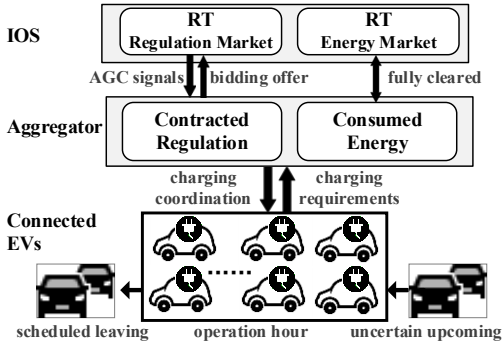

Figure 1. Illustrative schematic of EVA coordination between two sides.

The goal for EVA is to minimize its total cost in the wholesale market while sufficing all the charging requirements of EV owners. In the paper, the EVA is assumed to be a price-taker in the pool-based RT market. The aggregator submits quantity-only regulation capacity offers with predicted market prices. As such the regulation capacity offers at zero prices are often fully cleared.

The market setting of this paper is based on that of PJM, where all regulation needs are procured in its RT market. All regulation offers must be submitted prior to 14:15 day-ahead, however, to accurately reflect each resource's availability during the operating day, the regulation capability may be changed on an hourly basis up until 65 minutes prior to the start of the operating hour. In this paper, the RT regulation market allows an alteration of offered capacity one time-step (60 minutes) before the operation hour. Similar to PJM, Upward and downward regulation are treated as the same form of product. Thus, a resource must offer the regulation capacity in both directions. Additionally, there are two kinds of AGC signals in the PJM regulation market: the traditional regulation signal (RegA) for resources with low ramping capability, and the dynamic regulation signal (RegD) for resources with high ramp rate capabilities but limited energy availability. Without loss of generality, we only consider RegD signals here.

Under the performance-based regulation compensation scheme, PJM proposed performance scores (varying between 1 for perfect response and 0 for bad delivery), which reflect the actual regulation delivery with each 2-second instructed AGC signal. Note that it is difficult to explicitly calculate the performance score when a resource fails to follow AGC signals. In this paper, a perfect performance score is adopted in the mathematical model, and then actual regulation performance scores will be studied in the simulation.

## III. Mathematical Formulation

### A. Single EV Model

Firstly, the POP of the $i$th EV and its available regulation capacity at time $\tau$ are constrained through:

$$y_{i,\tau} \le x_{i,\tau} \le p_i^{max} - y_{i,\tau}, \quad \forall \tau \tag{1a}$$

$$y_{i,\tau} \ge 0, \quad \forall \tau \tag{1b}$$

where $x_{i,\tau}$, $y_{i,\tau}$ denote the POP and regulation capacity in MWh of the $i$th EV at time $\tau$, respectively; $p_i^{max}$ denotes the maximal charging power of the $i$th EV.

Notice that $\Delta E_i$ is used to denote the required charging energy, which can be calculated by the charging efficiency $\eta_i$, arrived SoC value $SoC_i^A$, required SoC value $SoC_i^E$ and rated energy capacity $E_i^{rated}$ of the $i$th EV:

$$\Delta E_i = \left(SoC_i^E - SoC_i^A\right) E_i^{rated} / \eta_i \tag{1c}$$

To ensure the charging requirements of EVs, POPs of the $i$th EV during the parking periods are subject to:

$$\sum_{\tau=t_i^{arv}}^{\tau=t_i^{dep}} x_{i,\tau} = \Delta E_i \tag{1d}$$

where $t_i^{arv}$, $t_i^{dep}$ denote the plug-in and plug-out times of the $i$th EV, respectively. Note that the assumption of hourly energy-neutral RegD is adopted in (1d), since RegD signals were originally designed to be energy neutral within 15 minutes[13]. In practice, energy deviations caused by actual regulation provisions will be corrected in the following *Part D*.

Next, an optimization problem is formulated to participate in both energy and regulation markets. Let $\lambda_\tau$, $\mu_\tau$ denote the energy clearing price and regulation clearing price at time $\tau$, respectively. To minimize the total cost of the $i$th EV while fulfilling its charging request, Problem (2) needs to be solved:

$$\min_{x_{i,\tau}, y_{i,\tau}} \sum_{\tau=t_i^{arv}}^{\tau=t_i^{dep}} \lambda_\tau x_{i,\tau} - \mu_\tau y_{i,\tau}$$

$$\text{s.t. (1a)-(1d)} \tag{2}$$

where $\mu_\tau = \mu_\tau^{rc} + \mu_\tau^{rp} m_\tau^{RgD}$; under performance-based regulation compensation, $\mu_\tau^{rc}, \mu_\tau^{rp}$ denote the regulation capacity and regulation performance clearing prices at time $\tau$, respectively; $m_\tau^{RgD}$ denotes the mileage at time $\tau$, which is calculated as the sum of absolute differences between two RegD signals.

Note that, the EVA has the same $\lambda_\tau$ and $\mu_\tau$ for each EV in the above optimization. In particular, optimal solutions of the problem (2), are denoted by $x_{i,\tau}^*$, $y_{i,\tau}^*$. Moreover, we define the integer charging flexibility index for the $i$th EV, denoted by $F_i = \lfloor 2 \cdot \Delta E_i / p_i^{max} \rfloor$.

*Proposition 1*: All EVs (where $i = 1, \cdots, N$) with the same $t_i^{arr}$, $t_i^{dep}$, $F_i$ values have the same optimal solutions to $\sum_i x_{i,\tau}^*$, $\sum_i y_{i,\tau}^*$ with an equivalent virtual EV who owns the same $t_i^{arr}$ and $t_i^{dep}$. And other parameters of this EV are as follows,

$$\Delta E = \sum_{i=1}^{N} \Delta E_i, \quad p^{max} = \sum_{i=1}^{N} p_i^{max}. \tag{3}$$

Please see detailed proof in the Appendix. With Proposition 1, EVA can drastically reduce variables in the optimizations.

*Remark 1*: Let $\rho$ denote the regulation compensation price in \$/MWh which EVA pays to EV owners for the profit share in the regulation market. Then the objective function of problem (2) changes as $\min_{x_{i,\tau}, y_{i,\tau}} [\sum_\tau \lambda_\tau x_{i,\tau} - (\mu_\tau - \rho) y_{i,\tau}]$. The EVA has the same $\rho$ for all EVs so that Proposition 1 still holds.

### B. Two-Stage Stochastic Strategy

Recall that the regulation capacity offer may be changed on an hourly basis one time-step before the operation hour $K$. As illustrated in Figure 2, EVA will solve the rolling-window optimization with $H$ look-ahead intervals at each time. At current time-step $K$, the MPC-based problem should submit the next time-step regulation capacity offer with the predicted market prices and information of upcoming EVs. Meanwhile, EVA modulates the RT charging power of connected EVs to respond to the cleared regulation capacity at time-step $K$.

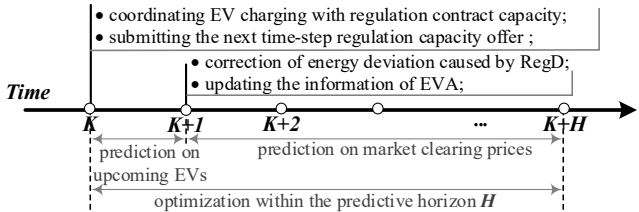

Figure 2. The MPC-based operation scheme.

Let $\mathcal{A}$ denote the set of all EVs connected to the aggregator. Information of set $\mathcal{A}$ is described by a 3-element matrix $[\Delta \mathbf{E}^\mathcal{A}, \mathbf{P}^{\mathcal{A},+}, \mathbf{T}^\mathcal{A}]$. The remaining charging energy, $\Delta \mathbf{E}^\mathcal{A}$, the maximal charging power, $\mathbf{P}^{\mathcal{A},+}$, the remaining parking periods, $\mathbf{T}^\mathcal{A}$, are deterministic input vectors of EVs in the set $\mathcal{A}$. We also model $\mathcal{F}$ to denote the set of all EVs which may arrive within the next hour. Note that, Proposition 1 is utilized to all upcoming EVs with respect to the departure deadlines and charging flexibility indexes. Therefore, the predicted input vectors of EVs in the set $\mathcal{F}$ are the remaining charging energy, $\Delta \mathbf{E}^\mathcal{F}$ and the maximal charging power, $\mathbf{P}^{\mathcal{F},+}$. Additionally, there is a finite number of equivalent virtual EVs in the set $\mathcal{F}$ so that the predicted input vectors of upcoming EVs have a finite dimension. A set of scenarios $\{s \in \Omega, \pi_s\}$ of the energy prices $\lambda_\tau^{(s)}$ regulation prices $\mu_\tau^{(s)}$ and parameters of upcoming EVs $[\Delta \mathbf{E}_s^\mathcal{F}, \mathbf{P}_s^{\mathcal{F},+}]$, are generated based on historical data and latest observations.

To minimize the total costs of EVA while managing all connected EVs to respond to the cleared regulation capacity at current time-step $K$, the following optimization problem needs to be solved:

$$\min Cost^{(s)} = \left( f_K + f_{K+1}^{(s)} + F^{(s)} + Pnlty^{(s)} \right) \tag{4}$$

where

$$f_K = \lambda_K \mathbf{1}^T \mathbf{X}_K^\mathcal{A} \tag{4a}$$

$$f_{K+1}^{(s)} = \lambda_{K+1}^{(s)} \mathbf{1}^T \left[ \mathbf{X}_{K+1}^{\mathcal{A},(s)}; \mathbf{X}_{K+1}^{\mathcal{F},(s)} \right] - \mu_{K+1}^{(s)} B_{K+1} \tag{4b}$$

$$F^{(s)} = \sum_{t=K+2}^{K+H} \left\{ \lambda_t^{(s)} \mathbf{1}^T \left[ \mathbf{X}_t^{\mathcal{A},(s)}; \mathbf{X}_t^{\mathcal{F},(s)} \right] - \mu_t^{(s)} \mathbf{1}^T \left[ \mathbf{Y}_t^{\mathcal{A},(s)}; \mathbf{Y}_t^{\mathcal{F},(s)} \right] \right\} \tag{4c}$$

$$Pnlty^{(s)} = \varphi \omega_K + \varphi' \omega_{K+1}^{(s)} \tag{4d}$$

s.t.

- Energy Constraints:

$$\mathbf{X}_K^\mathcal{A} + \sum_{t=K+1}^{K+H} \mathbf{X}_t^{\mathcal{A},(s)} = \Delta \mathbf{E}^\mathcal{A} \bullet \mathbf{g}(H, \mathbf{T}^\mathcal{A}) \tag{5a}$$

$$\sum_{t=K+1}^{K+H} \mathbf{X}_t^{\mathcal{F},(s)} = \Delta \mathbf{E}_s^\mathcal{F} \bullet \mathbf{g}(H-1, \mathbf{T}^\mathcal{F}) \tag{5b}$$

$$\left[ \mathbf{u}_t^\mathcal{A}; \mathbf{u}_t^\mathcal{F} \right] \circ \left[ \mathbf{X}_t^{\mathcal{A},(s)}; \mathbf{X}_t^{\mathcal{F},(s)} \right] = \mathbf{0}, \forall t = K+1, \ldots, K+H \tag{5c}$$

$$\mathbf{g}(H, \mathbf{T})^{(i)} = \min \left\{ 1, \frac{H}{\mathbf{T}^{(i)}} \right\}, \forall i \tag{5d}$$

- Operational Constraints:

$$\mathbf{Y}_K^\mathcal{A} \leq \mathbf{X}_K^\mathcal{A} \leq \mathbf{P}^{\mathcal{A},+} - \mathbf{Y}_K^\mathcal{A} \tag{6a}$$

$$\mathbf{Y}_t^{\mathcal{A},(s)} \leq \mathbf{X}_t^{\mathcal{A},(s)} \leq \mathbf{P}^{\mathcal{A},+} - \mathbf{Y}_t^{\mathcal{A},(s)}, \forall s, t = K+1, \ldots, K+H \tag{6b}$$

$$\mathbf{Y}_t^{\mathcal{F},(s)} \leq \mathbf{X}_t^{\mathcal{F},(s)} \leq \mathbf{P}_s^{\mathcal{F},+} - \mathbf{Y}_t^{\mathcal{F},(s)}, \forall s, t = K+1, \ldots, K+H \tag{6c}$$

$$\mathbf{Y}_t \geq \mathbf{0}, \forall t = K, \ldots, K+H \tag{6d}$$

$$\mathbf{1}^T \mathbf{Y}_K^\mathcal{A} \geq B_K - \omega_K \tag{6e}$$

$$\mathbf{1}^T \left[ \mathbf{Y}_{K+1}^{\mathcal{A},(s)}; \mathbf{Y}_{K+1}^{\mathcal{F},(s)} \right] \geq B_{K+1} - \omega_{K+1}^{(s)}, \forall s \tag{6f}$$

$$\omega_K, \omega_{K+1}^{(s)} \geq 0, \forall s \tag{6g}$$

where the first-stage variables are the current POP vector of the set $\mathcal{A}$, $\mathbf{X}_K^\mathcal{A}$, the current regulation capacity vector of the set $\mathcal{A}$, $\mathbf{Y}_K^\mathcal{A}$, the unfulfilled regulation capacity at current time-step $K$, $\omega_K$ and the next time-step regulation capacity offer, $B_{K+1}$; while the second-stage decision variables are $\mathbf{X}_t^{\mathcal{A},(s)}$, $\mathbf{X}_t^{\mathcal{F},(s)}$, $\mathbf{Y}_t^{\mathcal{A},(s)}$, $\mathbf{Y}_t^{\mathcal{F},(s)}$ and $\omega_{K+1}^{(s)}$ ($t \geq K+1$,), in which $s$ suggests the scenario index.

The aggregator objective (4) is the minimization of total operational costs within $H$ look-ahead intervals. The first term (4a) shows the EVA energy cost at operating time-step $k$. Although the aggregator can earn revenue from providing regulation, it should not take into account the current regulation payment when the regulation capacity has been already cleared. Note that EVA should submit a regulation capacity offer of the time-step $K + 1$ in the current operation. The next two components obtained from the EVA in both energy and regulation markets under s scenario are total costs at time-step $K + 1$, (4b), and total costs within remaining time-steps (4c). The penalty term for the aggregator over-capacity regulation bids can prevent the performance score degradation to some degree as expressed in (4d). The penalty factors of unfulfilled regulation capacities at time-step $K$ and $K + 1$ are denoted by $\varphi > 0$ and $\varphi' > 0$, respectively.

Constraints (5a)-(5b) model the charging requests of EVs in both set $\mathcal{A}$ and set $\mathcal{F}$. The element as expressed in (5d) denotes the ratio of required charging energy before departure deadline to that with $H$ look-ahead intervals. Furthermore, since the EV variables are expressed in the form of vectors, binary parameter vectors are defined by $\mathbf{u}_t^{\mathcal{A}}, \mathbf{u}_t^{\mathcal{F}}$ as the charging availability of EVs at time $t$ in the set $\mathcal{A}$ and set $\mathcal{F}$, respectively. When the element $\mathbf{u}_t^{\mathcal{A},(i)} = 1$, it means that the $i$th EV of set $\mathcal{A}$ is unconnected to the aggregator at time $t$. In (5c), operator ∘ is Hadamard product, also known as the element-wise product, which enforces POPs to be zero at the time-steps where EVs are unconnected. Constraints (6a)–(6d) reflect the operational limits of EVs. Note that the regulation services provided by the aggregator are the sum of regulation provisions offered by each connected EVs as expressed by (6e)–(6f). Parameter $B_K$ denotes the cleared regulation capacity at current operating time-step $K$. Here, we only consider the regulation capacity offers at time-step $K$ and $K + 1$ due to the RT regulation market settlement.

### C. Risk Aversion

In order to minimize EVA operational costs with the uncertainties of future arrival of EVs as well as energy and regulation prices, we incorporate a risk management by the use of CVaR. For the given confidence level denoted by $\alpha \in (0,1)$, Value at Risk (VaR) is defined as the cost in the $(1 - \alpha)$ fraction of worst-case outcomes, which is denoted by $z$ in this problem. CVaR computes the expected value of $z$, which can be mathematically defined as:

$$CVaR_{\alpha,s} = \min\left\{z + \frac{1}{1-\alpha}\pi_s\left\{\max\left\{Cost^{(s)} - z, 0\right\}\right\}\right\} \quad (7)$$

The CVaR can be incorporated into the optimization problem as:

$$\begin{array}{c} \underset{\substack{\mathbf{X}_K^{\mathcal{A}}, \mathbf{Y}_K^{\mathcal{A}}, B_{K+1}, \omega_K, \mathbf{X}_t^{\mathcal{A},(s)}, \\ \mathbf{X}_t^{\mathcal{F},(s)}, \mathbf{Y}_t^{\mathcal{A},(s)}, \mathbf{Y}_t^{\mathcal{F},(s)}, \omega_{K+1}^{(s)}}}{\text{minimize}} \left\{z + \alpha' \sum_s \pi_s v^{(s)}\right\} \end{array}$$

$$\text{subject to} \quad v^{(s)} \geq 0, \quad (8)$$
$$v^{(s)} \geq f_K + f_{K+1}^{(s)} + F^{(s)} + Pnlty^{(s)} - z,$$
$$(6a) - (6d), (7a) - (7d), (8a) - (8g),$$
$$\text{for each } s \in \Omega$$

where $\alpha' = (1 - \alpha)^{-1}$, and the linearization of (7) is achieved by adding new auxiliary variables $v^{(s)}$.

### D. Rolling-Horizon Operations

One can obtain optimal solutions $\mathbf{X}_K^{\mathcal{A}*}, \mathbf{Y}_K^{\mathcal{A}*}, B_{K+1}^*$ and $\omega_K^*$ of the first-stage variables after solving the problem (8). The optimal regulation capacity offer of the next time-step decided by the EVA is $B_{K+1}^*$, which should be submitted at current time-step K. For the $i$th EV, its regulation task denoted by $r_K^{(i)}$ can be allocated accordingly as:

$$r_K^{(i)} = \begin{cases} \mathbf{Y}_K^{\mathcal{A}*(i)} \times \dfrac{B_K}{\mathbf{1}^T \mathbf{Y}_K^{\mathcal{A}*}} & \omega_K^* = 0 \\ \mathbf{Y}_K^{\mathcal{A}*(i)} & \omega_K^* > 0 \end{cases} \quad (9a)$$

With each 2-second varying RegD signals $rg_D(\gamma, K) \in [-1,1]$ dispatched by ISO, the real-time charging power $chg_K^{(i)}$ of the $i$th EV can be calculated by,

$$chg_K^{(i)} = \mathbf{X}_K^{\mathcal{A}*(i)} - rg_D(\gamma, K) \times r_K^{(i)} \quad (9b)$$

As it is known, energy change of a plug-in EV comes from undertaking the regulation provision and performing the scheduled charging task. Therefore, in order to ensure the expected SoC level of the $i$th EV, the required charging energy $\Delta \mathbf{E}^{\mathcal{A},(i)}$ and remaining charging time $\mathbf{T}^{\mathcal{A},(i)}$ should be corrected after current time-step operation as,

$$\Delta \mathbf{E}^{\mathcal{A}(i)} = \Delta \mathbf{E}^{\mathcal{A}(i)} - \left(\mathbf{X}_K^{\mathcal{A}*(i)} - \sum_\gamma rg_D(\gamma, K) \times r_K^{(i)}\right), \quad (10)$$
$$\mathbf{T}^{\mathcal{A}(i)} = \mathbf{T}^{\mathcal{A}(i)} - 1$$

In addition, EVA should continuously manage the set $\mathcal{A}$ as EVs arrive and depart. Moreover, all predicted information needs to be updated based on latest observations at each time.

## IV. CASE STUDY

We evaluate the performance of the proposed methodology with a model of 1,000 EVs. Only home charging is considered. EVA participates in the RT energy and regulation markets while modulating the charging power of EVs plug-in during the operation. We take into account three kinds of charging patterns among the 1,000 heterogeneous EVs, whose parameters are generated based on the distribution of Table I. The historical market prices and regulation signals on Jul. 23, 2020, are used, and all data is available to the public in PJM[14]. Scenarios are generated by Monte Carlo method, where all predicted errors obey the Gaussian distributions.

TABLE I. PARAMETERS OF DIFFERENT TYPE EVs.

| Types | Num. | Arrival Time(h) | Departure Time(h) | Required △E(kWh) | Maximal charging rate(kW/h) |
|---|---|---|---|---|---|
| I | 600 | 16-23 | 6-13 | UD, [10,24] | UD, [4,8] |
| II | 200 | 0-7 | 14-21 | | |
| III | 200 | 8-15 | 22-5 | | |

*UD: uniform distribution.

Below are simulation results of the EVA operation problem:

Figure. 3 shows the optimal daily revenue versus predictive horizon size under three cases of different maximal charging rates. The impact of the horizon size is studied with the perfect prediction. It can be seen that the optimality gap will decrease with the larger horizon size, because more foreseen prices are used in the MPC optimization. Moreover, EVA attempts to provide more regulation capacity in the most lucrative time-steps, therefore, EVA daily revenue will increase with the larger maximal charging power of EVs.

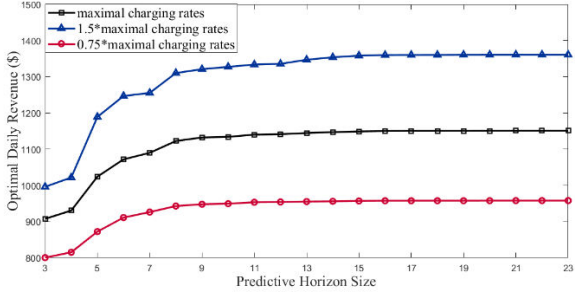

Figure 3. Regret Analysis under different maximal charging rates.

The performances under different confidence levels of CVaR, i.e., taking values of $\alpha$ from 0 to 1, are plotted in Figure 4. The predicted price error covariance denoted by $\varepsilon_p$ is also studied. We can find that the revenues keep dropping as $\alpha$ increases, clearly confirming the risk-averseness of regulation bidding decisions. And a higher covariance of predicted errors will lead to a poorer payment.

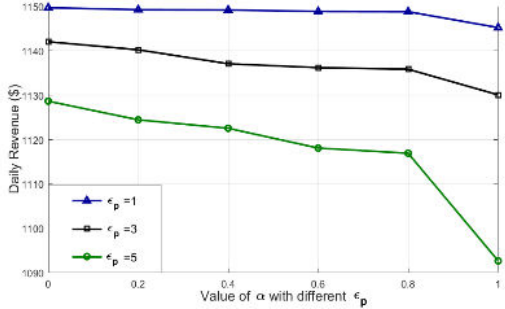

Figure 4. The CVaR versus confidence levels $\alpha$.

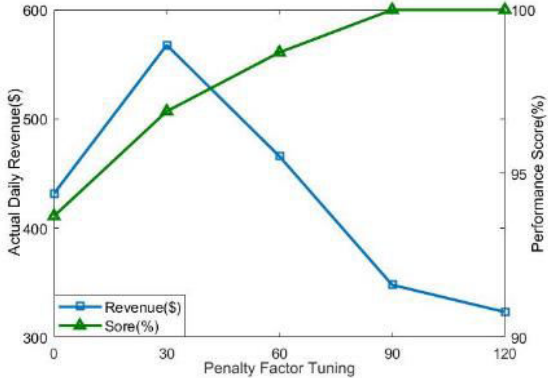

Figure 5. Performance score and actual revenue versus penalty $\varphi'$.

The daily EVA operation is simulated with the historical market prices and RegD signals in PJM. The parameters of whole problem are set to be $H=6$, $\varepsilon_p=3$, $\varepsilon_{EV}=5$, $\alpha=0.2$, $\varphi=115$. Figure 5 shows the impact of penalty factor tuning $\varphi'$ on the actual performance scores and according market revenues (using methods in [15]). Note that $\varphi'$ denotes the penalty factor caused by the over-capacity offer of next time-step. Compared to aforementioned results, the actual daily revenue in this case took a great tumble. This is because that the energy deviation of hourly RegD signals will cause the unexpected energy cost. Meanwhile, EV can only provide regulation when its POP>0 via unidirectional charging. From Figure 5, it can be seen that when $\varphi'$ increases, the actual revenue first increases and then decreases. It demonstrates the trade-off between regulation capacity and its performance score.

From EV owners' perspective, Table II gives the expected SoC deviations at departure of all EVs. Each EV's battery capacity is considered to be 20 kWh in this case. The result shows the worst SoC deviation may be acceptable in practice.

TABLE II. EXPECTED SOC DEVIATION AT DEPARTURE.

| △SOC(%) | (0,0.3) | (0.3,0.6) | (0.6,0.9) | (0.9,1.2) | (1.2,1.5) |
|---|---|---|---|---|---|
| Num. of EVs | 922 | 44 | 12 | 12 | 10 |

## V. CONCLUSION

In this paper, an MPC-based operation methodology for EVA participating in RT energy and regulation markets is proposed. Both the bidding strategy and charging allocation are considered in our formulation. The CVaR based two-stage stochastic programming is developed to address the uncertain upcoming EVs as well as energy and regulation prices. Simulations prove that the proposed MPC framework can effectively ensure the great delivery of cleared regulation and the fulfillment of EV charging requests. Numerical results also show the profitable revenue from the RT electricity market.

## APPENDIX

### A. Lemma and Proof

**Lemma 1:** There exists the optimal solution of problem (2), denoted by $(x_{i,\tau}^*, y_{i,\tau}^*)$ $\forall \tau$, which satisfies following conditions:

i) $x_{i,\tau}^*=0$, $y_{i,\tau}^*=0$;

ii) $x_{i,\tau}^*=p_i^{\max}$, $y_{i,\tau}^*=0$;

iii) $x_{i,\tau}^*=p_i^{\max}/2$, $y_{i,\tau}^*=p_i^{\max}/2$;

iv) there is at most one a time-step $\chi$, defined as marginal slot, where $(x_{i,\chi}^*, y_{i,\chi}^*) \notin \{(0,0),(p_i^{\max},0),(p_i^{\max}/2,p_i^{\max}/2)\}$.

**Proof:** Firstly, we make an equivalent transformation of problem (2) to prove $|x_{i,\tau}^* - p_i^{\max}/2| = |y_{i,\tau}^* - p_i^{\max}/2|$. Name $w_{i,\tau} = x_{i,\tau}^* - p_i^{\max}/2$; $z_{i,\tau} = p_i^{\max}/2 - y_{i,\tau}^*$. Then optimization (2) can be rewritten as,

$$\min_{w_{i,\tau}, z_{i,\tau}} \sum_{\tau \in T_i} \lambda_\tau w_{i,\tau} + \mu_\tau z_{i,\tau} + \frac{p_i^{\max}}{2}(\lambda_\tau - \mu_\tau)$$

$$\text{s.t.} \quad \sum_{\tau \in T_i} w_{i,\tau} = \Delta E_i - \frac{p_i^{\max}}{2}|T_i|, \quad (11)$$

$$|w_{i,\tau}| \leq z_{i,\tau}$$

$$0 \leq z_{i,\tau} \leq \frac{p_i^{\max}}{2}$$

where $T_i = [t_i^{arv}, t_i^{dep}]$; $|T_i|$ denotes the number of connected time slots.

Note that, equation $|w_{i,\tau}^*| = z_{i,\tau}^*$ holds for any $\tau \in T_i$ in (11). Because market prices will always be larger than zero in the objective function. Therefore, we can eliminate $z_{i,\tau}$ by using $|w_{i,\tau}|$ to replace it while assuring the optimality of the problem. Moreover, to simplify the parameters, we name $v_{i,\tau} = 2 \cdot w_{i,\tau}/$

$p_i^{max}$ as well as ignore the constant term and constant factor in the objective function. Then, we have,

$$\min_{v_{i,\tau}} \mathcal{V} = \sum_{\tau \in T_i} \lambda_\tau v_{i,\tau} + \mu_\tau |v_{i,\tau}|$$

$$\text{s.t.} \sum_{\tau \in T_i} v_{i,\tau} = \frac{2}{p_i^{max}} \Delta E_i - |T_i| = e_i \quad (\omega) \quad (12)$$

$$-1 \leq v_{i,\tau} \quad (\rho_\tau)$$

$$v_{i,\tau} \leq 1 \quad (\theta_\tau)$$

One can write the Karush-Kuhn-Tucker conditions for the minimization problem in (12) as

$$\sum_{\tau \in T_i} v_{i,\tau} = \frac{2}{p_i^{max}} \Delta E_i - |T_i| = e_i \quad (13a)$$

$$-1 \leq v_{i,\tau} \quad (13b)$$

$$v_{i,\tau} \leq 1 \quad (13c)$$

$$\frac{\partial \mathcal{V}}{\partial v_{i,\tau}} - \omega - \rho_\tau + \theta_\tau = 0 \quad \forall \tau \in T_i \quad (13d)$$

$$\omega \left( \sum_{\tau \in T_i} v_{i,\tau} - e_i \right) = 0 \quad (13e)$$

$$\rho_\tau (v_{i,\tau} + 1) = 0 \quad \forall \tau \in T_i \quad (13f)$$

$$\theta_\tau (v_{i,\tau} - 1) = 0 \quad \forall \tau \in T_i \quad (13g)$$

$$\rho_\tau \geq 0, \theta_\tau \geq 0 \quad \forall \tau \in T_i \quad (13h)$$

Since problem (12) has linear constraints, and the subgradient of nonlinear objective function can be solved as,

$$\frac{\partial \mathcal{V}}{\partial v_{i,\tau}} = \begin{cases} \lambda_\tau - \mu_\tau & \text{when } v_{i,\tau} < 0 \\ [\lambda_\tau - \mu_\tau, \lambda_\tau + \mu_\tau] & \text{when } v_{i,\tau} = 0 \\ \lambda_\tau + \mu_\tau & \text{when } v_{i,\tau} > 0 \end{cases} \quad (14)$$

Note that the equation $\rho_\tau \theta_\tau = 0 \ \forall \tau \in T_i$ holds, because the two constraints will not be active simultaneously as can be seen from KKT conditions (13f) (13g). Since $\omega$ is the optimal Lagrangian multiplier corresponding to the only equality constraint in (12). We separate our analysis into the following four cases:

1) for all time slots $\forall \tau \in T_i$ where $\lambda_\tau - \mu_\tau > \omega$, we have $\frac{\partial \mathcal{V}}{\partial v_{i,\tau}} - \omega = \rho_\tau - \theta_\tau > 0$. Then, $\rho_\tau > 0, \theta_\tau = 0$. From (13f),

$$v_{i,\tau}^* = -1; \Rightarrow x_{i,\tau}^* = 0, y_{i,\tau}^* = 0. \quad (15a)$$

2) for all time slots $\forall \tau \in T_i$ where $\lambda_\tau + \mu_\tau < \omega$, we have $\frac{\partial \mathcal{V}}{\partial v_{i,\tau}} - \omega = \rho_\tau - \theta_\tau < 0$. Then, $\rho_\tau = 0, \theta_\tau > 0$. From (13g),

$$v_{i,\tau}^* = 1; \Rightarrow x_{i,\tau}^* = p_i^{max}, y_{i,\tau}^* = 0. \quad (15b)$$

3) for all time slots $\forall \tau \in T_i$ where $\lambda_\tau - \mu_\tau < \omega < \lambda_\tau + \mu_\tau$, we have $v_{i,\tau}^* = 0; \Rightarrow x_{i,\tau}^* = p_i^{max}/2, y_{i,\tau}^* = p_i^{max}/2$. Otherwise, if $v_{i,\tau}^* > 0; \frac{\partial \mathcal{V}}{\partial v_{i,\tau}} - \omega > 0$. Then, we have $\rho_\tau > 0, \theta_\tau = 0$. $v_{i,\tau}^* = -1$, which is paradoxical; Else if $v_{i,\tau}^* < 0; \frac{\partial \mathcal{V}}{\partial v_{i,\tau}} - \omega < 0$. Then, we have $\rho_\tau = 0, \theta_\tau > 0$. $v_{i,\tau}^* = 1$, which is also paradoxical. To conclude, in this case, we have

$$v_{i,\tau}^* = 0; \Rightarrow x_{i,\tau}^* = p_i^{max}/2, y_{i,\tau}^* = p_i^{max}/2. \quad (15c)$$

4) for all time slots $\forall \tau \in T_i$ where $\lambda_\tau - \mu_\tau = \omega$, we have $v_{i,\tau}^* \in [-1,0]$; for all time slots $\forall \tau \in T_i$ where $\lambda_\tau + \mu_\tau = \omega$, we have $v_{i,\tau}^* \in [0,1]$. If there exists two time slots $(\alpha, \beta)$ in which $\lambda_\alpha - \mu_\alpha = \omega = \lambda_\beta - \mu_\beta$, or $\lambda_\alpha + \mu_\alpha = \omega = \lambda_\beta - \mu_\beta$, or $\lambda_\alpha + \mu_\alpha = \omega = \lambda_\beta + \mu_\beta$, we can find $v_{i,\alpha}^* + v_{i,\beta}^* = v'^*_{i,\alpha} + v'^*_{i,\beta}$, and both of them are optimal solutions. Therefore, we could always leave at most one time slot $v_{i,\chi}^*$ to be non-integer, and $\chi$ is defined as the marginal slot of the $i$th EV. □

### B. Proof of Proposition 1

**Proposition 1:** All EVs (where $i = 1, \cdots, N$) with the same $t_i^{arr}$, $t_i^{dep}$, $F_i$ values have the same optimal solutions to $\sum_i x_{i,\tau}^*$, $\sum_i y_{i,\tau}^*$ with an equivalent virtual EV who owns the same $t_i^{arr}$ and $t_i^{dep}$. And other parameters of this EV are as follows,

$$\Delta E = \sum_{i=1}^N \Delta E_i, p^{max} = \sum_{i=1}^N p_i^{max}.$$

**Proof:** Since all EVs have the same $t_i^{arr}$, $t_i^{dep}$, $F_i$ values, we use $t, F$ to denote the number of their connected time slots and integer charging flexibility index, respectively.

Firstly, we sort $2t+1$ elements of set $\{\lambda_\tau \pm \mu_\tau, -\infty\}, \forall \tau \in [t_i^{arv}, t_i^{dep}]$ to get a new ascending ordered set $\{O_{(1)}, ..., O_{(2t+1)}\}$. Next, we prove $\omega = O_{(F+1)}$ is the solution to the KKT conditions in (13a) – (13h), where $F$ is the integer charging flexibility index.

From discussions in Lemma 1, we group $t-1$ connected time slots (except marginal time slot $\chi$) into three cases:

1) There are $a$ time slots, where $v_\tau^* = -1$, and $\lambda_\tau - \mu_\tau \geq \omega$;
2) There are $b$ time slots, where $v_\tau^* = 1$, and $\lambda_\tau + \mu_\tau \leq \omega$;
3) There are $c$ time slots, where $v_\tau^* = 0$, and $\lambda_\tau + \mu_\tau \geq \omega$, $\omega \geq \lambda_\tau - \mu_\tau$;

For marginal time slot $\chi$, if $\omega = \lambda_\chi - \mu_\chi$ holds.

To conclude, following equations hold:

$$a + b + c = t - 1, \quad (16a)$$

$$1 + 0 \times a + 2 \times b + 1 \times c = F, \quad (16b)$$

$$(-1) \times a + 1 \times b + 0 \times c + v_\chi^* = e, \quad (16c)$$

where (16a) implies the total connected time slots; (16b) indicates the number of elements before $\omega = O_{(F+1)}$ in the ordered set; (16c) satisfy the energy constraint in optimization (12). To combine (16a) - (16c), we can get:

$$b - a = F - t, \quad (16d)$$

$$F - t + v_\chi^* = e. \Rightarrow v_\chi^* = e - F + t = e - \lceil e \rceil. \quad (16e)$$

Form (16e), we prove $v_\chi^* \in [-1,0]$ to be true. Thus, $\omega = O_{(F+1)}$ is feasible to the KKT constraints.

Else, when $\omega = \lambda_\chi + \mu_\chi$, (16a) and (16c) are still true, other equations change as:

$$1 + 0 \times a + 2 \times b + 1 \times c + 1 = F, \quad (16f)$$

$$b - a = F - t - 1, \quad (16g)$$

$$F - t - 1 + v_\chi^* = e. \Rightarrow v_\chi^* = e + 1 + t - F = e + 1 - \lceil e \rceil. \quad (16h)$$

Form (16h), we prove $v_\chi^* \in [0,1]$ to be true. Thus, $\omega = O_{(F+1)}$ is feasible to the KKT constraints.

To conclude, all EVs with the same $t_i^{arr}$, $t_i^{dep}$, $F_i$ values, have the same $\omega = O_{(F+1)}$ solution to its KKT conditions to (13). Briefly speaking, EVs will have the same charging preference (no-charging, half-maximal-charging, maximal-charging, marginal-charging) at any time slot.

Therefore, All EVs with the same $t_i^{arr}$, $t_i^{dep}$, $F_i$ values can be equivalent to a single EV while assuring the optimality.

□